\documentclass[aps,prd,endfloats*,showpacs]{revtex4}
\usepackage{graphicx,amssymb,amsfonts}
\newcommand{\be}{\begin{equation}}
\newcommand{\ee}{\end{equation}}
\newcommand{\ba}{\begin{eqnarray}}
\newcommand{\ea}{\end{eqnarray}}

\newcommand{\p}{\partial}
\newcommand{\del}{\triangledown}
\newcommand{\D}{\mathcal D}
\newcommand{\U}{\Upsilon}
\newcommand{\al}{\alpha}
\newcommand{\e}{\epsilon}
\newcommand{\tr}{{\rm tr}}
\newcommand{\np}{{Nuclear Phys.}}

\newcommand{\atmp}{Adv. Theor. Math. Phys.}

\begin{document}
\title{Topological QCD with a Twist}
\author{Alfred Tang}
\email{atang@alum.mit.edu}
\affiliation{Max Planck Instit\"ut f\"ur Kernphysik, 69120 Heidelberg,
Germany\\
and\\
The Chinese University of Hong Kong, Department of Physics, Shatin, Hong Kong.}
\date{\today}

\begin{abstract}
Non-supersymmetric Yang-Mill gauge theory in 4-dimension is shown to be dual
to 4-dimensional non-supersymmetric string theory in a twisted
$AdS^2(n)\times T_2$ spacetime background.  The partition function of a
generic hadron is calculated to illustrate the mathematical structure of the
twisted QCD topology.  Some physical implications of the twisted QCD topology
are discussed.
\end{abstract}
\maketitle

\section{Introduction}
Originally string theory was invented to explain hadron physics in the 70's.
It was thought from the beginning that there is a correspondance between
string theory and gauge theory~\cite{polyakov}.  Hadronic string theory was
eventually replaced by QCD in the 80's.  More recently interest in gauge-string
duality is revived when more precise statements about the duality between the
$AdS^5\times S_5$ superstring
theory and the ${\mathcal N}=4$ super-Yang-Mill gauge theory in the large
$N_c$ limit are made~\cite{maldacena,grubser,witten}.  Despite the amazing
advances in superstring theory, the
theoretical foundation of gauge-string duality is still not more
than a conjecture. Although supersymmetry is a beautiful
symmetry, it is not yet clear that it will make connection with the real world.
In low energy applications such as hadron physics, supersymmetry is assumed
to be irrelevant.  Given these considerations, statements of non-supersymmetric
gauge-string duality may find a niche in theoretical physics.  This work aims
to illustrate (but not prove) 4-dimensional gauge-string duality without
supersymmetry.  The language of a gauge theory spoken in this work is
mainly in QCD even though the arguments can be generalized to other types of
Yang-Mill gauge theories.  By the way of illustrating the relevance
of the present model, a sample calculation that approximates the partition
function of a generic hadron is given at the end.

\section{Gauge-String Duality}
The geometrical interpretation of gauge theory is quite old.  In lattice gauge
theory, a hadron is modelled by a torus with periodic and/or anti-periodic
boundary conditions.  The minimal coupling term in the total derivative
\be
D_\mu=\p_\mu+ig\,A_\mu
\ee
is analogous to the Christoffel symbol $\Gamma^a_{bc}$ in general relativity.
(In this paper, color indices are suppressed.  Lorentz indices are represented
by greek letters.  Roman indices stand for abstract indices in the context
of general relativity.)  The correspondance
between the gauge field and the Christoffel symbol is easily derived as
\be
A_\mu\to{1\over i4g}\,\Gamma^\nu_{\mu\nu}.
\label{amu}
\ee
In general relativity, Christoffel symobels are contracted with all
components of a vector
being derived as in $\del_a t^b=\p_a t^b+\Gamma^b_{ac}t^c$.  A covariant
derivative generally involves the entire volume around a point.
Eq.~(\ref{amu}) on other hand suggests that a total derivative in Yang-Mill
theory depends on
only 2 components, $D_\mu t_\nu=\p_\mu t_\nu+igA_\mu t_\nu$.  If $D_\mu$ is
interpreted as a translation operator on a 1-form $t_\nu$, the motion will
be generated on a 2-dimensional plane defined by the $\mu,\nu$ indices.  It
means that any given Yang-Mill gauge configuration has a stratified topology
by default.  So far
the theoretical foundation of a worldsheet description of planar Yang-Mill
theory was simply assumed~\cite{thorn}.  Although the present argument is not
a proof of the existence of planar Yang-Mill theory, it hopes to offer
additional justifications for the worldsheet approach of gauge theory.

\begin{figure}
\includegraphics[scale=0.5]{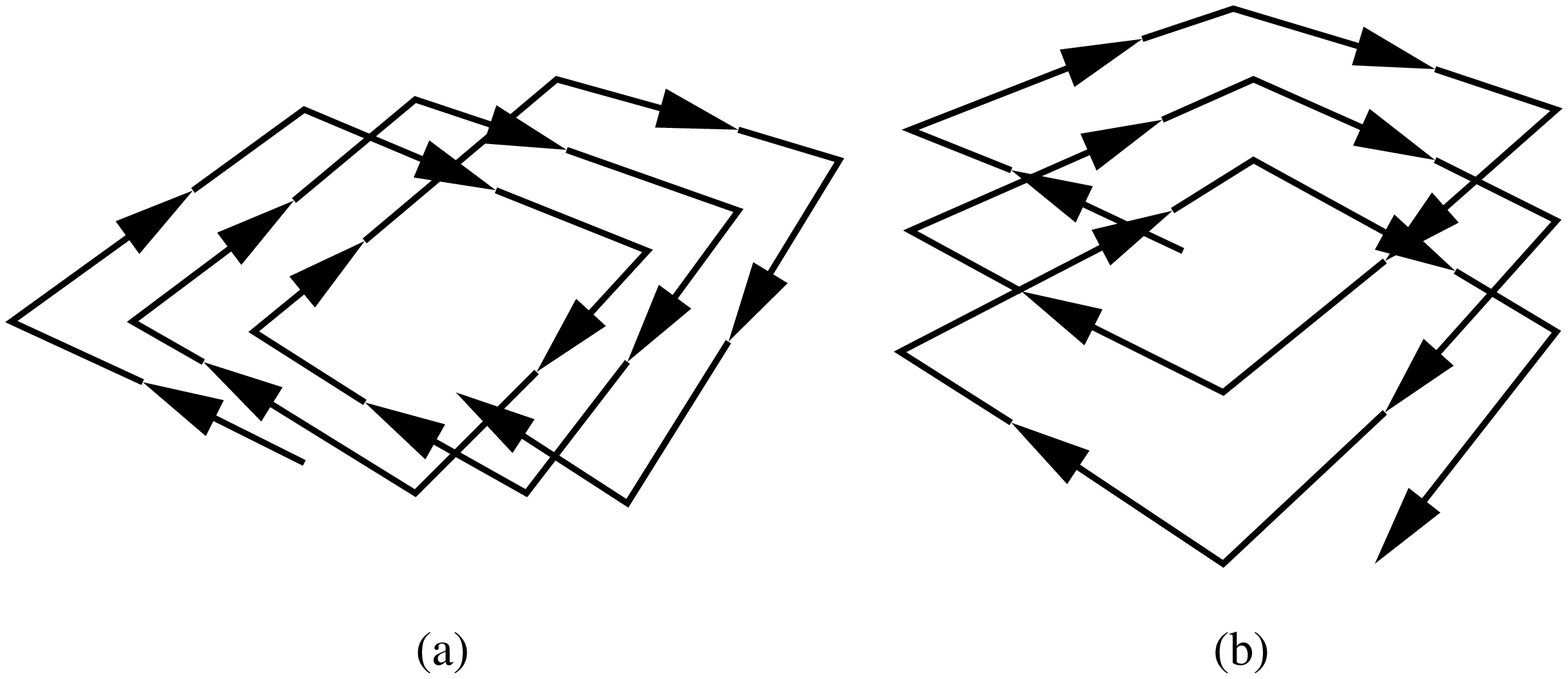}
\caption{\label{torsion}
The non-communitative properties of the displacement operatior,
$D_\mu=\p_\mu+igA_\mu$ can be interpreted as (a) curvature and
(b) torsion.  The arrows represent the spacetime translation generated by
$D_\mu$.  Curvature causes the loops to translate laterally as shown in
(a).  Torsion is signified by the presence of a spiral translation of the
loops as in (b).}
\end{figure}

\begin{figure}
\includegraphics[scale=0.4]{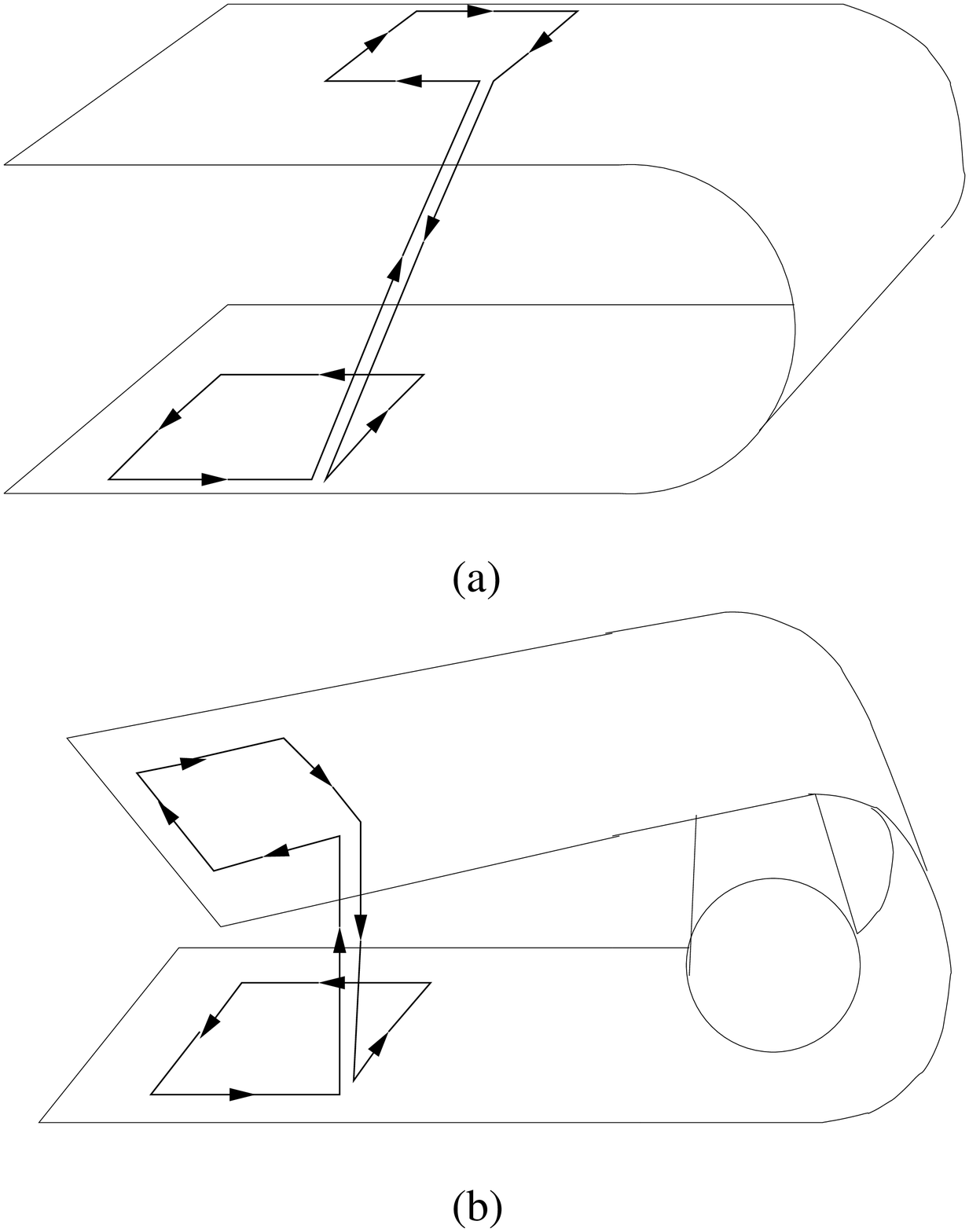}
\caption{\label{topo}
The spacetime displacement generated by the loop $[D_\mu,D_\nu]$ on a
2-dimensional surface creates a residual vector called the ``structure vector''
reperesenting curvature and torsion pointing away from the surface.  Sets of
loops and counter loops
on 2 surfaces cancel the equal and opposite structure vectors as shown in
(a).  The components of the structure vectors parallel to the generating loops
can be eliminated effectively by twisting the surfaces as shown in (b).
The cylinder in (b) provides counter loops to cancel the structure vectors
of the embedding outer surface at the edge.}
\end{figure}

A loop on a 2-dimensional spacetime hyper-surface can be generated by the
commutation bracket
\be
[D_\mu,D_\nu]=ig(\p_\mu A_\nu-\p_\nu A_\mu)-g^2[A_\mu,A_\nu],
\label{comm}
\ee
where $\mu\neq\nu$.  Non-commutative geometry
suggests that spacetime has structures.  Figure~\ref{torsion} illustrates
two kinds of topological structures (curvature and torsion) traced out
by displacement vectors.  Curvature is a measure of the geometrical features
of the worldsheet while torsion measures the rotational motion perpendicular
to a tower of worldsheets.  The first term on the right hand side of
Eq.~(\ref{comm}) involves derivatives that rotates.  Rotational vector
is perpendicular to the plane such that it is taken to represent torsion.  The
second term involves a series of Christoffel symbols connecting points around
a loop so that it is taken to represent the curvature of spacetime.  Together
the two terms on the right hand side of Eq.~(\ref{comm}) represent the
parallel (curvature) and perpendicular (torsion) components of a so-called 
``structure vector'' generated by the
commutation bracket on the left hand side and is pointing away from the
worldsheet.  On a carefully chosen surface, two equal and opposite structure
vectors generated by a loop and a counter-loop are cancelled as shown
in Figure~\ref{topo}(a).  The parallel component of the structure vector
can in principle be gauged away by twisting the the surface as shown in
Figure~\ref{topo}(b).  In retrospect, it is understood that any arbitrary
manipulation of surfaces can always be performed because the string action
will be integrated over all random Riemann surfaces using the path-integral
method.  The
twist degree of freedom is limited at the edge of a folded surface so that a
cylinder is introduced underneath the folded surface to provide a means to
generate counter-loops to cancel the structure vectors of the outer surface
(see Figure~\ref{topo}(b)).  The gauge degrees of freedom cannot be illustrated
in a spacetime diagram.  The structure of worldsheet in Figure~\ref{topo}
implicitly defines the gauge field configuration by correspondance based on
the principle of gauge-string duality.
The parallel component of the structure vector has a geometric interpretation
of a force generated by the gauge field.  Its removal on the twisted surface
implies that the surface is equi-potential.  Torsion is a unique feature of the
non-Abelian nature of the gauge field.  The cancellation of the perpendicular
components of the structure vectors means that a non-Abelian theory is
projected onto
an Abelian theory.  Although this conclusion is a surprise, it is consistent
with observations in lattice gauge calculations where the full $SU(2)$ string
tension is shown to be calculable in a maximally Abelian $U(1)$
gauge~\cite{hart}.  The size of the outer twisted surface is limited by the
embedded toroidal surface.  The smallest outer surface cannot overlap with the
largest inner surface.  It means that both the inner and outer
surfaces have fuzzy supports and a fuzzy thickness $L^2$.  The support
of the worldsheets effectively defines the physical locations of the gauge
fields.
Since the inner and outer gauge surfaces are disjoint, the gauge fields around
the two surfaces are also disjoint by default.

Na\"ively Eq.~(\ref{comm}) may be taken to be the same as $F_{\mu\nu}=0$
because both curvature and torsion are gauged away by the twisted topology when
$\mu\neq\nu$.  For $\mu=\nu$, $F_{\mu\nu}=0$ is true only in
untwisted space when the loop collapses into a pair of lines running in
opposite directions in a flat background.  In twisted QCD topology, the pair
of lines is separated by the twist.  A possible parameterization of the
stress tensor is
\ba
F_{\mu\nu}&\equiv&{1\over2}\,g^2[\,\delta^2(\Sigma_i)+
\delta^2(\Sigma_o)]\,\delta_{\mu\nu},\label{f1}\\
F^{\mu\nu}&\equiv&{1\over2}\,g^2{1\over L^2}\,\delta^{\mu\nu},\label{f2}
\ea
where $\delta^2(\Sigma)$ is a 2-dimensional delta-function restricting the
the separation of the line pair to the surface $\Sigma$.  The inner and
outer equi-potential gauge surfaces are labelled as $\Sigma_i$ and
$\Sigma_o$ respectively.  $\delta_{\mu\nu}$
and $\delta^{\mu\nu}$ are Kronecker delta-functions.  The choice of the
forms of Eqs.~(\ref{f1}) and (\ref{f2}) essentially picks a gauge for the
induced metric $h_{\mu\nu}$ on the twisted string worldsheet.

\begin{figure}
\includegraphics[scale=0.6]{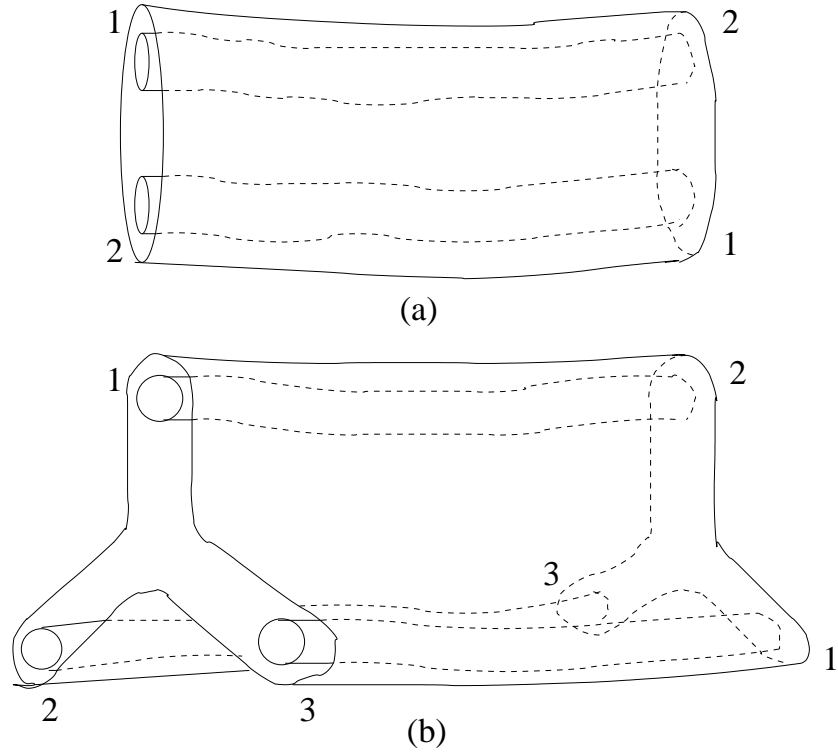}
\caption{\label{hadron}
Sketches of the topologies of (a) a meson and (b) a baryon.  The numbers
show the identification of the boundaries.  The topology of the gauge
field of a hadron consists of a twisted torus with negative curvature and $n$
edges (2 edges for a meson and 3 for a baryon and so on) called $AdS^2(n)$ and
an embedded torus $T_2$.  There are two possible twists in the case of the
baryon---one corresponding to a twist of $120^\circ$ and another to
$240^\circ$.
The identification in part (b) corresponds to a twist of $120^\circ$.}
\end{figure}

\begin{figure}
\includegraphics[scale=0.6]{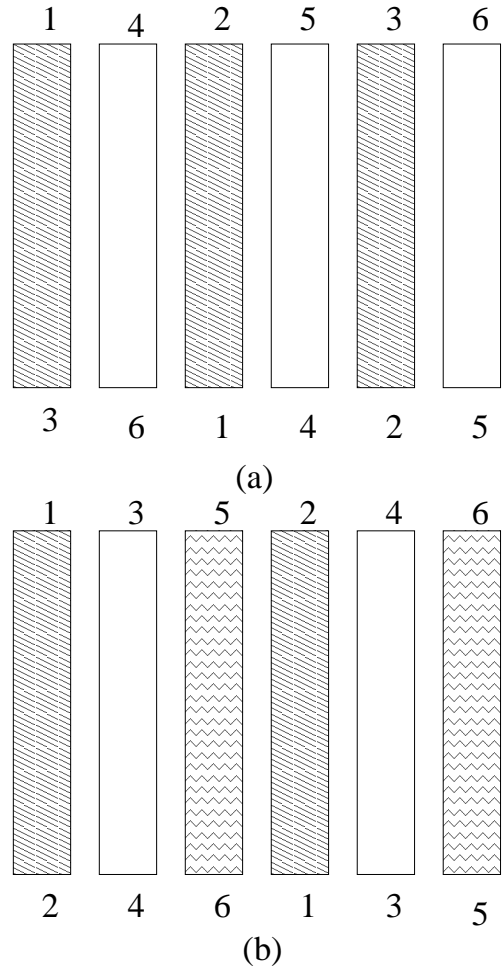}
\caption{\label{fact}
The factorizations of a 6-quark bound state $(n=6)$.  There are two possible
factorizations depending on the different identifications of the
boundaries---(a) a twist of $120^\circ$ resulting in 2 loops of 3 coils
each and (b) a twist of $240^\circ$ giving 3 loops of 2 coils each.  Each
bar represents an edge of the twisted torus and corresponds to one coil.}
\end{figure}

The combination of the twist,
periodical boundary conditions in space and time, and the number of edges
on the closed string worldsheet (taken
to be the same as the number of constituent quarks) completes the picture of
the global structure of the gauge field as shown in Figure~\ref{hadron}.
The topology of the gauge field of a hadron consists
of a twisted torus with negative curvature and $n$ edges $AdS^2(n)$ and
an embedded torus $T_2$.  It is assumed that the matter fields of the
constituent quarks live inside the torus $T_2$ from symmetry considerations.
The
gauge-string topology of a meson corresponds to a twisted torus with 2 edges
($n=2$).  The closed
mesonic worldsheet is not a M\"obius strip because the surface is orientable.
The baryonic topology ($n=3$) has a special name called the ``triniton.''  The
twisted topology is analogous to the M\"obius strip in that multiple strips
are joined into one around the loop.  Since all the edges are one, it is
reasonable to assume that there is only one embedded torus $T_2$.  This
condition restricts $n$ to prime numbers.  As shown in Fig.~\ref{fact}, if
$n=6$, the identification
of the twisted topology makes possible 2 embedded tori with 3 coils each or 3
embedded tori with 2 coils each.  Only prime $n$ will prevent a factorization
of the embedded torus $T_2$.  This restriction may offer a possible explanation
for the experimental observation that only pentaquark is observed, but not
tetraquark.

\section{Twisted Partition Function}
The partition function of the string theory in an $AdS^2(n)\times T_2$
background is calculated in this section as an example to illustrate the
mathematical structure of the twisted QCD topology.  It will soon be
obvious that there are still many unsolved mathematical problems so that
reasonable approximations are made along the way.  In order to simplify
the discussion, the Yang-Mill gauge action $S_g$ contains only the free gauge
term,
\be
S_g={1\over4}\int d^4x\,\tr\, F^{\mu\nu}F_{\mu\nu}.
\ee
The partition function in gauge theory is defined as
\be
Z_g=N\int\D A\,e^{-S_g}\to\int\D A_i\D A_o\,e^{-S_g}
\label{zg}
\ee
$N$ is a normalization constant.  Numerical constants will be absorbed into
$N$ at various stages of the calculation.  For the sake of simplicity, the
presence of the normalization constant is understood implicitly so that $N$
will be dropped explicitly from now on.  $A_o$ and $A_i$ are the gauge fields
around the $AdS^2(n)$ and $T_2$ surfaces respectively.  The factorization
$\D A\to\D A_i \D A_o$ is chosen because the gauge fields on the inner and
outer surfaces are considered disjoint in an effective sense.  This choice
also makes it possible to factorize the partition function to yield
simple results at a later point.  By choosing
the lightcone gauge $(A^+=0)$ in the Faddeev-Popov procedure, the ghost fields
are eliminated
\be
Z_g=\int\D A_i\D A_o\,\delta(A^+)\,e^{-S_g}\label{z}
\ee
so that the string worldsheet topology of the gauge field remains simple in
the discussion to follow.  Substitution of Eqs.~(\ref{f1}) and (\ref{f2}) into
Eq.~(\ref{z}) shows that the partition function can be factorized as
\be
Z_g=Z_i Z_o,\label{zg2}
\ee
where
\be
Z_{\{i,o\}}=\int\D A_{\{i,o\}}\,\exp\left(-{g^4 f_c\over4L^2}\int
d\Sigma_{\{i,o\}}\right).\label{zio}
\ee
$f_c$ is the color factor calculable from the structure constants of
the Yang-Mill algebra.  Let $h$ be the induced metric on the worldsheet,
$x$ the displacement vector and $R$ the radius of the hadron.  The Jacobian in
$\D A\to J\,\D h\,\D x$ is approximated through dimensional analysis as
\be
J\sim{1\over gR^2}.\label{j}
\ee
The coefficient of the action in Eq.~(\ref{zio}) is interpreted
as the string tension.  It can be absorbed into $x$ by a change of variable.
At last Eq.~(\ref{zio}) is transformed as
\be
Z_{\{i,o\}}={L\over g^3R^2\sqrt{f_c}}
\int\D h_{\{i,o\}}\,\D x\,\exp\left(-{1\over2}\int d\Sigma_{\{i,o\}}\right),
\label{zs}
\ee
Eq.~(\ref{zs}) has
the form of a string partition function.  A path-integral over Yang-Mill
gauge configurations is now transformed to a path-integral over string
worldsheets.  This procedure is similar to a Penrose transform.

In topological QCD, both $AdS^2(n)$ and $T_2$ are
genus 1 surfaces.  The former differs from the latter by a twist and a
negative curvature.  Curvature is a local property in twisted QCD topology.
A path-integral sums over all the topological equivalent surfaces globally.
Genus 1 surfaces of both signs of
local curvature are topologically equivalent and are therefore
indistinguishable to the path-integral.  The induced matric on an untwisted
space $h_{\mu\nu}$ is related to that on a twisted space $\tilde{h}_{\mu\nu}$
by a twist function $Z_n$.
It is also reasonable to assume that the area of the string worldsheet is
unaffected by the twist so that the action is invariant in the twist degree of
freedom.  This way the partition function of the twisted outer surface $Z_o$ is
related to that of the untwisted inner surface by the twist function $Z_n$
as in
\be
Z_o=Z_n Z_i.\label{zt}
\ee
With Eq.~(\ref{zt}), Eq.~(\ref{zg2}) can be rewritten as
\be
Z_g=Z_n Z^2_i.
\ee
Typically the Polyakov version of the string action
\be
S_p={1\over2}\int d\sigma d\tau \sqrt{-g}\,g^{ab}\p_a x^\mu \p_b x_\mu
\ee
is used in the worldsheet path-integral
\be
Z_i=\int\D g_{ab}\D x^{\mu}\,e^{-S_p}.
\label{zT}
\ee
The evaluation of Eq.~(\ref{zT}) for bosonic string in the critical
dimension $(D=26)$ is well-known~\cite{polchinski} and is given as
\be
Z_c={N\over2}\int_{D(\Gamma)}{d^2\tau\over2\pi\tau^2_2}\,
e^{4\pi\tau_2}(2\pi\tau_2)^{-12}
\left|\prod^\infty_{n=1}1-e^{2i\pi n\tau}\right|^{-48},
\label{ztorus}
\ee
with $\tau\equiv\tau_1+i\tau_2$. Let $\mathcal U$ be the upper half of
the complex plane, then
\be
D(\Gamma)= \left\{\tau\in{\mathcal U} \,\left|\,{\rm Re}\,\tau\leq{1\over2},
\,|\tau|\geq 1 \right.\right\}.
\ee
In non-critical dimensions, conformal anomaly contributes an extra factor
$Z_L$ so that $Z_i=Z_L\,Z_c$ and
\be
Z_L=\int\D\phi\,\exp\left(-{26-D\over48\pi}\,S_L\right).
\label{zL}
\ee
The Liouville action $S_L$ has very complicated mathematical
structures~\cite{aldrovandi} and is still an unsolved problem today.
Nevertheless approximate solutions of the Liouville field theory have been
constructed based on reasonable guesses~\cite{dorn}.  The approximate
contribution to $Z_i$ from the Liouville action is given as
\be
Z_L\simeq\left[\left(26-D\over48\right)
\mu\,b^{-2b^2}{\Gamma(1+b^2)\over\Gamma(1-b^2)}
\right]^{Q/b}{\U_0\over\U(-Q)}.
\label{zL2}
\ee
(See Appendix~\ref{liouville}.)  In the standard model, $D=4$.  The twist
function $Z_n$ of a spin-1 field is approximated from results in
reference~\cite{russo} as
\be
Z_n=-e^{-i\pi\e}2i\sin\pi\e\prod^\infty_{n=1}\left(1-e^{2i\pi\e}k^n\right)
\left(1-e^{-2i\pi\e}k^n\right),
\label{zn}
\ee
where $k=\exp(2i\pi\tau)$ and $\tau$ is the period matrix in the
$\Theta$-function.  The parameter $\e=m/n$ where
$m\in\{\mathbb{Z}\,|\, 0<m<n\}$ measures the twist.  At last the combination
of Eqs.~(\ref{ztorus}), (\ref{zL2}) and (\ref{zn}) give the final expression
of the partition function
\be
Z_g=Z_n Z_L^2 Z_c^2\sim{L^2\over g^6 R^4 f_c}\,Z_n.
\label{ans}
\ee
The last part of Eq.~(\ref{ans}) highlights the dependencies of $Z_g$.

\section{Discussions and Conclusion}
Non-supersymmetric Yang-Mill gauge theory in 4-dimension is shown to be dual
to non-supersymmetric 4-dimensional string theory in an $AdS^2(n)\times T_2$
background.
The topology of the gauge field represented by spacetime worldsheet has
a twisted structure.  On any given time slice, the topology shows multiple
gauge surfaces with negative curvature.  These multiple surfaces trace out
a single surface globally in a way analogous to a M\"obius strip.  The $T^2$
torus embedded underneath the twisted outer surface
along the edge is thought to contain the matter field of the constituent quark.
In the
example of a baryon, the triunity structure of the gauge field suggests the
possibility that there is only one global quark field that manifests itself
as 3 separate matter fields locally.  In this case, the quark field couples
to itself via back-reaction.  If mulitple constituent quarks are simply
mirror images of a
single quark, confinement is automatically obtained because a quark cannot
deconfine from itself.  Since the purpose of this work is to illustrate
the concept of twisted QCD topology in the simplest manner, matter fields are
excluded from this anaylsis in order to focus solely on the dynamics of
the gauge field.  However, in order to have
meaningful discussions on hadron phenomenology, matter fields and interaction
temrs must be included into a realistic action.  Nevertheless it is encouraging
to see that confinement emerges naturally in the present model without any
need to tweak the action.  The coupling of the loops and counter-loops via the
structure vectors may provide a mechanism for color superconductivity.  The
subjects of matter fields, the interaction terms, color superconductivity,
confinement and asymptotic freedom will be discussed in future works.  In the
present work, the partition function of a generic
hadron is approximated in Eq.~(\ref{ans}) as a sample calculation
to illustrate the mathematical structure of twisted QCD topology
which is still work in progress.  It is expected that the final
phenomenologically meaningful results will take on slightly different forms
as the theory becomes more mature.  The solutions of
many difficult mathematical problems are still wanting and a more
mathematically rigorous proof of gauge-string duality is still in need,
possibly in the language of twister algebra and Penrose transform.
Future works will test the validity of twisted QCD topology by calculating
physical observables such as Regge trajectories, generalized parton
distributions and electromagnetic form factors $G_E/G_M$ to be compared with
experimental data.  The present work predicts the non-existence of
tetraquark and hexaquark as opposed to the prediciton of their existence by
standard model calculations~\cite{vijande}.  Future experiments of exotic
bound states with either positive or negative discoveries will provide
further insights into the structure of hadrons.

\begin{acknowledgments}
This work was supported in part by the Max Planck Instit\"ut Exchange Scholar
Program at Heidelberg and the Postdoctoral Fellowship Program in the Department
of Physics at the Chinese University of Hong Kong.  The author thanks J.
Friedman for a discussion on the concept of torsion and gauge theory.  He also
thanks J. Maldacena for taking his precious time to read the first draft of the
manuscript and for offering his helpful comments.

\end{acknowledgments}

\appendix
\section{Liouville Action}\label{liouville}
The Liouville action $S_L$ is defined as
\be
S_L={1\over4\pi}\int d^2x\left[(\p_a\phi)^2+4\pi\mu\,e^{2b\phi}\right],
\label{liou}
\ee
where $\mu$ is called the cosmological constant and $b$ is a dimensionless
coupling constant.  Approximate solutions of the 3-point function of the
Liouville field theory have been constructed~\cite{dorn}.  An example is
\ba
C(\al_1,\al_2,\al_3)&=&\int\D\phi\,e^{-S_L}
V_{\al_1}(x_1)V_{\al_2}(x_2)V_{\al_3}(x_3)\\
&=&\left[\pi\mu b^{-2b^2}{\Gamma(1+b^2)\over\Gamma(1-b^2)}
\right]^{(Q-\sum\al_i)/b}\nonumber\\
&&\quad\times{\U_0\U(2\al_1)\U(2\al_2)\U(2\al_3)\over
\U(\al_1+\al_2+\al_3-Q)\U(\al_1+\al_2-\al_3)\U(\al_2+\al_3-\al_1)
\U(\al_3+\al_1-\al_2)}.\label{c}
\ea
The special function $\U(x)$ is defined as
\be
\log\U(x)=\int^\infty_0{dt\over t}\left[\left({Q\over2}-x\right)^2e^{-t}-
{\sinh^2\left({Q\over2}-x\right){t\over2}\over
\sinh{bt\over2}\sinh{t\over2b}}\right],
\ee
with
\be
Q=b+{1\over b}.
\ee
Given that
\be
\U_0=\left.{d\U(x)\over dx}\right|_{x=0},
\ee
and
\be
V_\al(x)=e^{2\al\phi(x)},
\ee
it can be seen that $Z_L$ can almost be obtained from Eq.~(\ref{c}) by
setting $\al_1=\al_2=\al_3=0$ if it were not the factor of
$(D-26)/48\pi$ in Eq.~(\ref{zL}).  It turns out that the factor can be absorbed
into the cosmological constant $\mu$ and the field variable $\phi$ of the
Liouville action $S_L$ in
Eq.~(\ref{liou}) by a change of variables.
At last an approximate solution of $Z_L$ is
estimated from the 3-point function as given by Eq.~(\ref{zL2}).
The variable $b$ in
Eq.~(\ref{zL2}) is changed to
\be
b\to\sqrt{{26-D\over48\pi}}\,b.
\ee
It is noted that $Z_n\neq 1$ as $D\to26$.  The reason is that Eq.~(\ref{zL2})
is an approximation based on an educated guess and is valid for
non-critical dimensions only.

\end{document}